\documentclass[11pt]{article}
\usepackage{moriond,epsfig}

\def\be{\begin{equation}}
\def\ee{\end{equation}}
\def\bea{\begin{eqnarray}}
\def\eea{\end{eqnarray}}

\begin{document}
\vspace*{4cm}
\title{New Views into High Redshift Star-formation from GOODS} 

\author{ R. Chary }

\address{Spitzer Science Center, MS220-6, Caltech, Pasadena, CA 91125; rchary@caltech.edu}

\maketitle\abstracts{
The GOODS-North field centered around the historical Hubble Deep Field-North provides one of the
richest multiwavelength datasets compiled, spanning radio to X-ray frequencies, for studying the
formation and evolution of galaxies at high redshift. In particular, the unprecedented
sensitivity of the Spitzer/GOODS 24$\mu$m observations 
allows an unbiased measure of dust-enshrouded star-formation and AGN activity
in typical L$_{*}$ galaxies rather than the extreme,
hyperluminous galaxies which far-infrared surveys detect. We consider a 
spectroscopically selected sample of 226 galaxies
at z$>$1.5 of which 135 are
in the redshift range 1.5$<z<3$. 
Less than 1/4 of galaxies considered here are detected at 24$\mu$m. 
They primarily have red UV slopes but consitute about only 20\% of the red galaxies
in this sample
indicating that the red UV colors of the majority of objects is due to an evolved stellar population.
Although 24$\mu$m sources are sparse in number, their combined energy output in the 1.5$<z<3$ range
exceeds the combined UV luminosity of the sample by a factor of $\sim$30. 
We also find that AGN, identified by their X-ray to infrared luminosity ratios, account for $<$10\% of the sources
considered and contribute less than 30\% of the total budget in this redshift range. 
The infrared luminous galaxies which are increasingly being found
at high-z appear to dominate the global energetics of the Universe out to z$\sim$3 as has previously been predicted.
}

\section{Scientific Motivation} 
The redshift range 1$<z<$3, corresponding to a cosmic time of 2$-$6 Gyr after the Big Bang, encompasses various
interesting observational characteristics of galaxy evolution. This includes the peak in the comoving number density of quasars,
the dramatic growth in the stellar mass of galaxies and the $\sim$(1+z)$^{4}$ evolution in the comoving number density 
of infrared luminous galaxies as detected by ISOCAM at 15$\mu$m and SCUBA at 850$\mu$m. The ultraviolet and infrared
pictures of galaxy evolution have not been entirely consistent views over this redshift range with discrepancies arising
from the uncertain correction for dust extinction at short wavelengths to the inability of mid- and far-infrared surveys
to detect galaxies at the faint end of the luminosity function. After estimates of these corrections have been made,
the agreement in the comoving bolometric luminosity density is around a factor of $2-3$. This obfuscates the fact that estimates
of star-formation, AGN activity and stellar mass in individual galaxies do not agree to within an order of magnitude making it
difficult to pinpoint the physical trigger for starburst or quasar activity in individual galaxies, the duration of such 
activity, the relative growth of supermassive black hole mass and stellar mass, and the relative contribution of nucleosynthesis and
accretion to the cosmic energy budget over cosmic time.

The Great Observatories Origins Deep Survey (GOODS) has targetted an area of 165 arcmin$^{2}$ each around the Hubble Deep Field-North
and the Hubble Ultra Deep Field in the South with the Hubble/ACS in the BV$iz$ passbands (Giavalisco et al. 2004)
and with Spitzer using the IRAC and MIPS instruments
in the 3.6$-$24$\mu$m range (Dickinson et al. 2005; in preparation). Together 
with the deep Chandra X-ray data (Alexander et al. 2003), radio observations 
with the VLA and ground-based imaging and spectroscopy from 
ESO, NOAO, Keck and Subaru, the two GOODS-fields yield one of the deepest, most homogeneous, publicly available data sets compiled
to study the different facets of galaxy evolution over a wide redshift range.

I present results from a multiwavelength study of the properties of 226 galaxies in the GOODS-N field
with spectroscopic redshifts greater than 1.5. The spectroscopic selection is the primary selection effect
introduced into these results
since specific objects were targetted for their pre-{\it Spitzer} characteristics. For example, the sample
includes Chandra X-ray sources (Barger et al. 2002), SCUBA 850$\mu$m sources (Chapman et al. 2005)
and objects detected in the deep VLA 1.4 GHz survey (Richards 2000 and Morrison et al. 2005). In 
addition, the vast majority of redshifts
were obtained from the Treasury Keck Redshift Survey (TKRS; Wirth et al. 2004) which targetted a magnitude limited sample
with R$_{\rm AB}<24.4$~mag. Broadly speaking, this biases the results presented here towards AGN and strong emission line
objects which can be alleviated in the future as spectroscopic redshifts for the entire GOODS sample becomes 
available.

\section{Properties of the Sample and AGN fraction}

The galaxies selected here have spectroscopic redshifts of $z>1.5$. Their apparent magnitudes
span the $27>z_{AB}>20$ mag range. The 24$\mu$m detected sources have a median $z_{{\rm AB}}$ of 23.8 mag, about
0.8 mags brighter than the median of the entire sample. X-ray sources account for 12.8\% (29/286) of the sample.
The surface density of X-ray sources averaged over the entire
GOODS-N field is about 1:16 relative to MIPS 24$\mu$m sources 
and 1:110 relative to 3.6$\mu$m selected sources at GOODS depths. The relatively high X-ray fraction here is 
mostly attributable to the spectroscopic bias and partly to the greater X-ray sensitivity and
thereby higher X-ray source density in the central region of the GOODS field.

To determine the contribution of AGN to the luminosity of the sources considered here, 
we calculated the ratio of infrared (8-1000$\mu$m)
luminosity to X-ray luminosity, L$_{IR}$/L$_{X}$ and compared
this with the X-ray photon index ($\Gamma$). The infrared luminosity is derived from the 24$\mu$m
flux and redshift using two approaches. One is to use the library of 
Chary\&Elbaz (2001) and Dale\&Helou (2002)
template SEDs of star-forming galaxies
and identify the template which provides the closest match to the 24$\mu$m flux at the redshift
of the source. The other is to use a typical obscured AGN SED i.e. Mrk231 and scale it to fit
the 24$\mu$m flux of the source. Typically, since the SED is flatter in $\nu$F$_{\nu}$ for an 
AGN the bolometric correction from the Mrk231 SED is smaller 
than that for the starburst SEDs resulting in L$_{IR}$ values which differ by a factor of $\sim$7.
A significant source of uncertainty in the redshift range under discussion is the presence
of silicate absorption at rest-frame 9.7$\mu$m which would suppress the observed 24$\mu$m flux
when the source is in the redshift range $1.1<z<1.9$. Since the template SEDs do not have 
any strong intrinsic silicate absorption built in, they will underestimate the bolometric luminosity
of those sources if silicate absorption is present.

For AGN which might be X-ray undetected because of Compton-thick column densities, we assess
the presence of an AGN from the IRAC colors of the sources (e.g. Lacy et al. 2004). We do 
not expect to find such sources in this sample because of the spectroscopic selection effects. 
We find that application of the IRAC color-color cut defined in Lacy et al. (2004) or Stern et al. (2005)
results in the selection of 31 non X-ray detected sources and 13 X-ray detected sources. 
57\% of the
non X-ray sources are detected in the 24$\mu$m while if they were mostly Compton-thick AGN, they
should all have been 24$\mu$m detected. All but 9/31 have derived L$_{IR}$/L$_{X}$ limits that are $>$100.
The spectroscopic redshifts of these source are in the range $z\sim2.5\pm0.5$. 
In this redshift range, for typical AGN luminosities,
the X-ray column densities should be $>10^{24}$~cm$^{-2}$ for the sources to be undetected in the Chandra
data and even the most optimistic models for the hard X-ray background indicate that such sources contribute
at most 20\% of the background.
Thus, we conclude that the IRAC color-color AGN diagnostics derived from shallow surveys, when applied
to deeper IRAC datasets, have a high rate of contamination from distant starburst galaxies as the 
1.6$\mu$m peak of the stellar SED moves through the IRAC passbands. 

As a result, to derive the AGN fraction in our sample, we consider AGN to be sources which are hard X-ray detected and with
L$_{IR}$/L$_{X}$ values $<100$. We find that of the 29 X-ray detected sources, only about 12 (5.3\%) of these are 
AGN dominated with X-ray luminosities in the range 10$^{43-45}$ ergs/s. Objects which have 
100$<$L$_{IR}$/L$_{X}<$1000 are assumed to be transition objects (5.8\%) with increasing contribution from 
star-formation while all the remaining objects (89\%) are assumed to be starburst dominated. 

We find the total contribution to the UV light from AGN and transition objects
in this sample is 3$\times$10$^{12}$~L$_{\odot}$. After applying a correction based on the UV-slope, this value
is enhanced to 1.6$\times$10$^{14}$~L$_{\odot}$ while the derived L$_{IR}$ for the 24$\mu$m detected objects is
8$\times$10$^{13}$~L$_{\odot}$. The contribution from AGN is 45\% while transition objects
account for 55\%. The dust correction factor to the UV luminosity for the AGN in this sample is 
therefore about $\sim$30 and in reasonable agreement between the 24$\mu$m and UV-slope estimates.

\section{High-redshift Starbursts}

\subsection{Star-formation Rates}

We measure the star-formation rates of the majority of objects which are starburst dominated
using the ultraviolet continuum (L$_{UV}$) and
the derived infrared luminosity (L$_{IR}$) from the 24$\mu$m flux. In the 8 cases where the source
also has an 850$\mu$m detection, the L$_{IR}$ was calculated by simultaneously fitting the 24$\mu$m
and 850$\mu$m flux with the template SEDs. We derive an extinction correction
from the ultraviolet slope using the prescription of Meurer et al. (1999) for comparison with
the observed infrared emission. We find that 24$\mu$m sources are preferentially redder in their E(B-V)
colors but that not all red sources are detected at 24$\mu$m. The median E(B-V) colors derived
from the UV slope for galaxies at 1.5$<z<3$
is 0.18 mag while those of the 24$\mu$m detected starburst population is 0.32 mag. Of the galaxies
which are redder than the median, only about 20\% of the objects are detected at 24$\mu$m. This seems
to suggest that on an individual galaxy basis, the UV-slope technique seems to overpredict the
dust correction, most likely a result of the age-extinction
degeneracy whereby the steep UV-slope of an evolved stellar population is attributed to reddening.

To assess if the UV-slope accurately measures the dust correction in the reddened objects, we
plot the corrected star-formation rate derived from the UV and the infrared (Figure 2). We find that for typical
infrared luminous galaxies with L$_{IR}\sim10^{11}$~L$_{\odot}$, the agreement between the UV-slope
corrected SFR and the 24$\mu$m derived SFR is very good. However, as we move towards more luminous
systems which probably are powered by optically thick starbursts, there is a systematic divergence
in the sense that the UV-slope underestimates the bolometric luminosity and thereby the star-formation rate.

In terms of global averages, the total UV luminosity of these objects is 3.6$\times$10$^{12}$~L$_{\odot}$, while
the integrated IR luminosity of the 24$\mu$m detected sources, which constitute 20\% of the starbursts,
is 7.6$\times$10$^{13}$~L$_{\odot}$. This implies that the 
dust correction factor for this sample is about 20 while the correction from the UV-slope is
more than an order of magnitude larger. 
Including the transition objects classified from the previous section as star-forming objects
boost the luminosity from the 24$\mu$m detected sources to 1.2$\times$10$^{14}$~L$_{\odot}$.
It should be emphasized that the large correction factors stated here are not a correction to the 
co-moving UV-luminosity
density in this redshift regime, which can be no larger than a factor of 3 (Chary\&Elbaz 2001), but rather 
an illustration of the difficulties associated with deriving global SFR from UV-slope
corrected densities in this sample of objects. 

\subsection{Physical Properties}

We used the multiband photometry afforded by GOODS to measure the physical properties of the galaxies
as a function of their star-formation rate. We fit Bruzual-Charlot SEDs to the photometry, using
a salpeter IMF, and solar metallicity template while varying the age of the starburst, mass and internal
extinction.
We find that there is no apparent correlation between the A$_{V}$ derived from the template fits and
the detection of 24$\mu$m flux from the objects. However, we find that infrared luminous starbursts
appear to have higher specific star-formation rates than normal galaxies (Figure 3). This is possibly an evolutionary effect
in that galaxies having higher stellar masses preferentially occur in massive dark matter halos where the distribution
of gas is more strongly concentrated towards the nucleus. This results in a a larger frequency of optically thick
star-forming regions which reprocess the bulk of the UV flux into the far-infrared. As a result, there is practically no
change in the observed UV slope (which arises from optically thin regions) with increasing far-infrared flux.

\section{Conclusions}

Using the luminosity calculated here, we can estimate the contribution from AGN and starbursts to the total energy budget
at $1.5<z<3$. We find for our sample, the AGN contribution to the total luminosity density
is about 20-30\%, depending on the fraction of transition objects
that are classified as starbursts or AGN. This is likely to be an upper limit due to the bias in this spectroscopic sample.
For the starbursts, we find that the slope of the UV-continuum provides a reasonable indicator as to which galaxies are dusty.
However, only about 20\% of red galaxies in this sample are detected at 24$\mu$m implying that a red UV-slope does
not necessitate the presence of dust. Star-formation rates derived from the UV continuum, are almost always underestimates
but corrections derived from the UV-slope are inaccurate at ULIG luminosities. Furthermore, because of the age-extinction
degeneracy, the UV-slope tends to overcorrect the SFR for individual galaxies. Only about 25\% of galaxies in this sample
show thermal dust emission but their combined luminosity exceeds that of all UV-selected field galaxies implying that
dust plays a dominant role in high-z star-formation.

\section*{Acknowledgements}
This paper would not be possible without key contributions from various members of the GOODS team who are gratefully 
acknowledged. 
Support for this work, part of the Spitzer Space Telescope Legacy Science Program, was provided by NASA through an award
issued to the Jet Propulsion Laboratory, California Institute of Technology under NASA contract 1407.

\begin{figure}
\psfig{figure=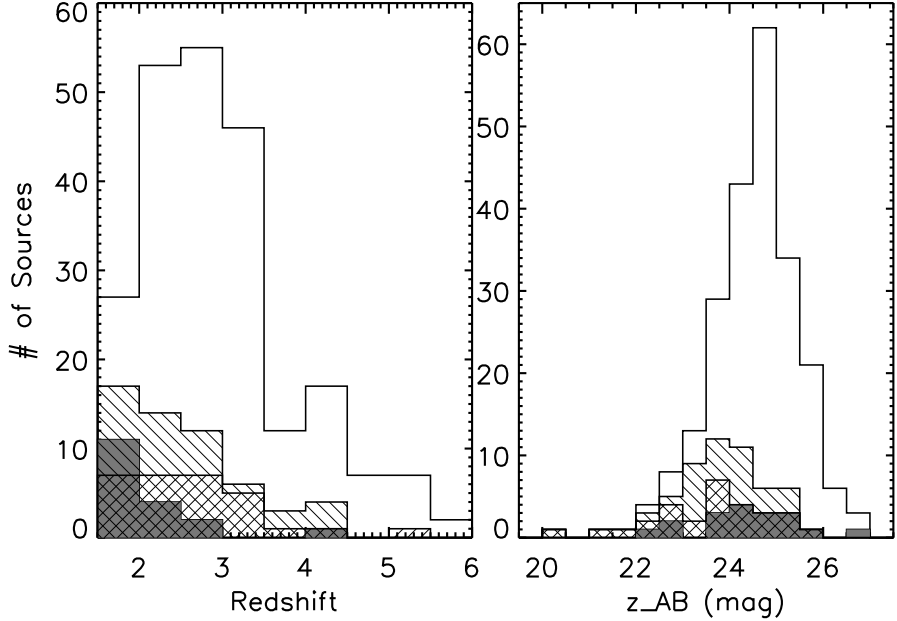,height=3.5in}
\caption{Redshift and $z-$band apparent magnitude
distribution of the sample of sources considered. The histogram is the distribution of all sources,
the backward hashed region represents MIPS 24$\mu$m sources, the forward hatched region represents X-ray sources
while the grey solid region indicates radio sources. The sample is biased towards a high detection rate of X-ray sources
since the spectroscopic follow-up was first initiated on X-ray detected sources. In this sample, 3.5\% of
sources are SCUBA 850$\mu$m detected,
8.0\% are 1.4 GHz detected, 12.8\% are X-ray detected while 24.8\% are 24$\mu$m detected.
\label{fig:figone}}
\end{figure}

\begin{figure}
\psfig{figure=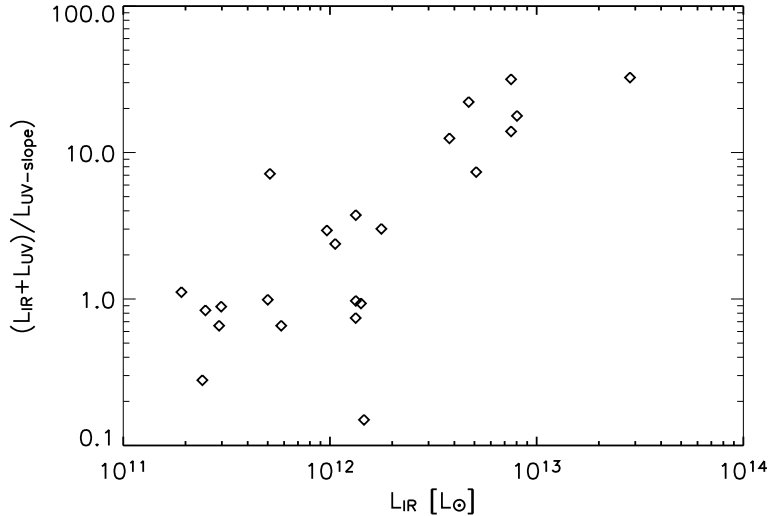,height=3in}
\caption{
Comparison between SFR estimates from the extinction corrected UV-slope technique and the 24$\mu$m flux
for star-forming galaxies at 1.5$<$z$<$3.
The agreement between the different estimates in infrared luminous galaxies with L$_{IR}\sim10^{11}$~L$_{\odot}$,
corresponding to SFR of 17~M$_{\odot}$/yr, 
is very good. However, the different estimates diverge
towards larger L$_{IR}$ galaxies where the SFR is probably in the optically thick regime where the fraction
of UV photons escaping is negligibly small. It is possible that the L$_{{\rm IR}}$ estimates for some of the objects
with L$_{{\rm IR}}>10^{12.5}$~L$_{\odot}$ have been overestimated due to unusually strong PAH emission. 
\label{fig:figtwo}}
\end{figure}

\begin{figure}
\psfig{figure=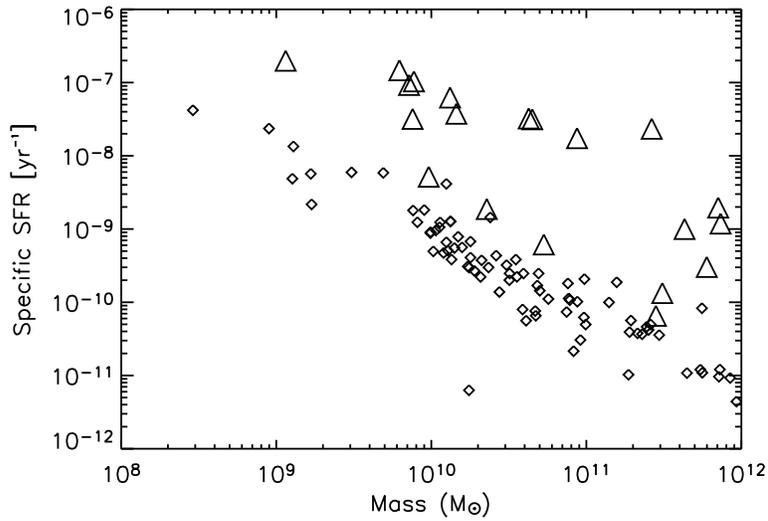,height=3in}
\caption{
Comparison between the SFR/Stellar mass (specific star-formation rate) for 24$\mu$m detected starburst galaxies and the
galaxies in the spectroscopic sample. Triangles indicate the 24$\mu$m detected galaxies whose SFR
has been derived from L$_{IR}$. Diamonds are the 24$\mu$m undetected galaxies whose SFR is from the extinction-uncorrected UV flux.
24$\mu$m detected galaxies have a higher specific SFR than normal galaxies and also appear to be more massive than the field galaxies.
\label{fig:figthree}}
\end{figure}

\end{document}